\documentclass[10pt]{iopart} 
\usepackage{amsfonts,amssymb,iopams}
\usepackage{epsfig}
\usepackage{iopams,cite}
\usepackage{graphicx}
\usepackage{xcolor,soul}
\usepackage[normalem]{ulem}

\newcommand{\text}[1]{\mbox{\scriptsize{#1}}}
\usepackage[utf8]{inputenc}

\expandafter\let\csname equation*\endcsname\relax
\expandafter\let\csname endequation*\endcsname\relax
\usepackage{amsmath}

\begin{document}

\title{Sub-cellular mRNA localization modulates the regulation of gene expression by small RNAs in bacteria}
\author{Hamid Teimouri$^{1}$, Elgin Korkmazhan$^{2}$, Joel Stavans$^{3}$ and Erel Levine$^{1,4}$}
\address{$^1$Department of Physics, Harvard University, Cambridge, MA, 02138
\\
$^2$Harvard College, Cambridge, MA, 02138
\\
$^3$Department of Physics of Complex Systems, Weizmann Institute of Science, Rehovot 76100, Israel\\
$^4$FAS Center for Systems Biology, Harvard University, Cambridge, MA, 02138}
\eads{\mailto{elevine@fas.harvard.edu}}

\begin{abstract}
Small non-coding RNAs can exert significant regulatory activity on gene expression in bacteria. In recent years, substantial progress has been made in understanding bacterial gene expression by sRNAs. However, 
recent findings that demonstrate that families of mRNAs show non-trivial sub-cellular distributions raise the question of how localization may affect the regulatory activity of sRNAs. Here we address this question within a simple mathematical model. We show that the non-uniform spatial distributions of mRNA can alter the threshold-linear response that characterizes sRNAs that act stoichiometrically, and modulate the hierarchy among targets co-regulated by the same sRNA. We also identify conditions where the sub-cellular organization of cofactors in the sRNA pathway can induce spatial heterogeneity on sRNA targets. Our results suggest that under certain conditions, interpretation and modeling of natural and synthetic gene regulatory circuits need to take into account the spatial organization of the transcripts of participating genes. 
\end{abstract}

\pacs{}

\newpage
A class of non-coding RNAs, known as small RNAs, play a crucial role in the regulation of gene expression in bacteria \cite{frohlich2009activation,waters2009regulatory,storz2011regulation}. The regulatory effect of the sRNA is achieved through base pairing interactions with target mRNAs that lead to modulation of translation and mRNA stability \cite{Aiba2007134,gottesman_micros_2005,storz_controlling_2004,prevost_small_2007}. For the best-studied class of small RNAs, this interaction depends on the RNA chaperone Hfq and involves the degradosome ribonuclease RNase E \cite{morita2011rnase}. 
In some cases, these interactions also lead to prolonged sequestration or enhanced degradation of the sRNA itself \cite{masse2003coupled}. This non-catalytic nature of sRNA-mRNA interaction can bring about unique regulatory behaviors such as a threshold-linear response to regulatory signals and robust noise repression\cite{levine_quantitative_2007,levine_small_2007,levine2008small}. Often, a single sRNA can specifically regulate multiple targets\cite{jost2013regulating}. Non-catalytic interactions between the sRNA and at least some of its targets lead to a built-in hierarchy and regulatory cross-talk among them\cite{levine_quantitative_2007,shimoni2007regulation,mitarai_efficient_2007,jost2013regulating,fei_determination_2015}. In particular, it has been hypothesized that the threshold-linear response, by which sRNA transcription sets a threshold for target expression, could be important for bacteria to fight random fluctuations and transient signals\cite{levine2008small}. In recent years, these features and others have been described in detail, both theoretically and experimentally \cite{jost_small_2011,feng_qrr_2015,arbel2016transcript,Kumar2016Frequency}.

Recent advances in super-resolution microscopy and fluorescence in situ hybridization (FISH) allow for direct measurement of the spatial arrangement of macromolecules {\it in vivo}\cite{montero_llopis_spatial_2010,gahlmann2014exploring,buxbaum2015right,fei_determination_2015,moffitt2016spatial,kocaoglu2016progress,govindarajan2016things,jan2016finding}. As a result of such  efforts it has been experimentally shown that the mRNAs of some genes can be localized to specific regions in the cell \cite{keiler2011rna,nevo2012subcellular,buskila_rna_2014,weng2014spatial}. In particular, mRNAs  that encode inner membrane bound proteins are enriched in the vicinity of the membrane\cite{kawamoto_implication_2005,fei_determination_2015,moffitt2016spatial}. 
Spatial distribution of mRNAs  and small non-coding RNAs in the bacterial cell can have two main consequences. First, localization of mRNAs in specific parts of the cell may be convenient for synthesis of locally functioning proteins, including membrane binding proteins. Second, spatial localization of mRNA could affect their life-cycle, including their stability, translation, and regulation. In particular, localization of transcripts may make them more or less accessible to sRNAs \cite{kawamoto_implication_2005}, or modulate the mechanisms by which sRNAs regulate their targets.

Here we use a simple mathematical model to investigate the possible implications  of sub-cellular localization on post-transcriptional regulation by sRNAs, focusing in particular on the effect of mRNA distribution on key characteristics of sRNA regulation. Our model predicts that localization of mRNAs can have  quantitative  effects on sRNA-based gene regulation, without changing its functional properties. When comparing the regulatory effect of an sRNA on its multiple targets, localization can modulate the hierarchy established among these targets.  We also investigate the effect of spatial localization of co-factors of the sRNA pathway on the regulation of their targets. Finally, we discuss implications of spatial localization to modeling of genetic circuits in the context of both systems and synthetic biology.  


\section{Theoretical Method}
Rod-shaped bacteria such as {\it{Escherichia}} and {\it{Pseudomonas}}, regardless of their detailed morphologies, can be idealized as circular cylinders. For the sake of simplicity we assume that the spatial distributions of an sRNA and its targets are longitudinally symmetric, and   perform our analysis by considering a cross section of the cell as a disk of radius $R$. 

The mRNAs of genes that encode membrane bound proteins are enriched near the cell membrane. 
This is probably due to co-translational insertion of membrane proteins. In this mechanism, nascent membrane proteins may be targeted to the membrane once a signal recognition peptide or a membrane-binding domain has been synthesized, even before completion of translation of the entire protein. 
Insertion of a nascent peptide to the membrane brings the translating ribosome, and with it the mRNA and other bound proteins, to the vicinity of the membrane. In our model we assume that an mRNA can either be in a state in which it is anchored to the membrane and resides in its vicinity, or in a freely diffusing state. Due to the fast diffusion in the bacterial cell, we do not consider the precise position of the mRNA in the cytosol.

Thus, we separate the cell interior into two regions, the cytosol (an inner disk with radius $R_{1}$, referred to as region 1) and the membrane vicinity (the remaining part extending from $R_{1}$ to $R$, referred to as region 2). The transition rate $d^{(12)}$ from region 1 to region 2 is governed by diffusion, and (to first order) is related with the diffusion constant $D$ through $d^{(12)} = 6\pi D$. The transition rate $d^{(21)}$ is related with the rate at which mRNAs stop being tethered to the membrane, and depends on the structure of the proteins they encode and on the rate of translation \cite{Elgin2017}. Transcripts that do not encode co-translationally inserted proteins translocate symmetrically between the two regions.  

Recent findings suggest that the RNA chaperone Hfq is enriched near the membrane, although the mechanism behind that is not known \cite{diestra2009cellular}. Since some of small RNAs have high affinity to Hfq, it is possible that the interaction with Hfq increases the affinity of some sRNAs to the membrane region. We therefore investigate both the case where the small RNA  diffuses freely between regions,
$d_s^{(12)} = d_s^{(21)}$, as well as the case where the rate $d_s^{(21)}$ is associated with unbinding from Hfq. 

Our model accounts for a single sRNA and its $n$ targets. sRNAs are transcribed with rate $\alpha_s$ and mRNAs of species $i$ with rate $\alpha_i$. These RNAs are degraded with rates $\beta_s$ and $\beta_i$, respectively. Since our focus is on sRNA-mediated degradation, we ignore spatial dependence of these degradation rates (although some ribonucleases are known to be associated with the membrane \cite{khemici_rnase_2008,moffitt2016spatial,Chen2015Genome}). 
The stoichiometric degradation between the sRNA and target $i$ is described by an interaction strength $k_{i}^{(j)}$. The membrane association of Hfq and RNase E, which is involved in sRNA-mediated degradation of target mRNAs, suggests that this rate could be different in the two regions of the cell, indicated by the superscript $j$. 

Out interest is in the steady-state concentrations $s^{(j)}$ of the sRNA and $m^{(j)}_{i}$ of the mRNAs of species $i$ in region $j$. We let $V_{1}=\pi R_{1}^{2}$ and $V_{2}=\pi(R^{2}-R_{1}^{2})$ be the volumes of the two regions, and set $\gamma = V_2/(V_1+V_2)$. With these,  the average concentration of mRNA of species $i$ is  given by 
$m_{i}=(1-\gamma) m_{i}^{(1)} + \gamma m_{i}^{(2)}$ and that of the sRNA by $s=(1-\gamma) s^{(1)}+ \gamma s^{(2)}$.

The dynamics of the model is given by the mass-action equations
\begin{eqnarray}\label{srna}
\frac{ds^{(1)}}{dt}=\alpha_{s} - \beta_{s}s^{(1)} - \sum\limits_{i=1}^{n}k^{(1)}_{i}m^{(1)}_{i}s^{(1)} + \frac{d_{s}^{(21)}s^{(2)}}{V_{1}}-\frac{d_{s}^{(12)}s^{(1)}}{V_{1}}\\
\frac{ds^{(2)}}{dt}=\alpha_{s}-\beta_{s}s^{(2)}- \sum\limits_{i=1}^{n}k^{(2)}_{i}m^{(2)}_{i}s^{(2)} +\frac{d_{s}^{(12)}s^{(1)}}{V_{2}}-\frac{d_{s}^{(21)}s^{(2)}}{V_{2}}\nonumber
\end{eqnarray}
for the sRNA, and
\begin{eqnarray}\label{targeti}
\frac{dm^{(1)}_{i}}{dt}=\alpha_{i}-\beta_{i}m^{(1)}_{i}-k^{(1)}_{i}m^{(1)}_{i}s^{(1)}_{i}+\frac{d_{i}^{(21)}m^{(2)}_{i}}{V_{1}} -\frac{d_{i}^{(12)}m^{(1)}_{i}}{V_{1}}\\
\frac{dm^{(2)}_{i}}{dt}=\alpha_{i} - \beta_{i}m^{(2)}_{i} - k^{(2)}_{i}m^{(2)}_{i}s^{(2)}_{i} + \frac{d_{i}^{(12)}m^{(1)}_{i}}{V_{2}} -\frac{d_{i}^{(21)}m^{(2)}_{i}}{V_{2}}\nonumber
\end{eqnarray}
for the mRNA of species $i$. The steady state concentrations are obtained by setting all time derivatives to zero and solving these equations. In general it is not feasible to obtain the full analytical solution for this system of non-linear equations, but a robust numerical solution is easily obtained. In all calculations we use a set of parameters that has been experimentally verified in \textit{E. coli} (Table 1).


\begin{table}
\begin{center}
\begin{tabular}{llr}
\hline
\multicolumn{2}{c}{Model Parameters} \\
\cline{1-3}
Parameter    & Meaning & Estimated Value \\
\hline
$\alpha_{m}$   & Transcription rate of target mRNA &  1 nM/min      \\
$\alpha_{s}$   & Transcription rate of  sRNA &  1 nM/min \\
$\beta_{m}$    & Degradation rate of target mRNA     &  1/10 min$^{-1}$     \\
$\beta_{s}$    & Degradation rate of sRNA    & 1/50 min$^{-1}$     \\
$k$   & Interaction strength of sRNA and mRNA  & 1/50 (nM min)$^{-1}$   \\
$R$   & Radius of the cell  & $1$ $\mu m$   \\
$d_{m}^{(12)}$   & Transition rate of mRNAs towards membrane & $10$ ${\mu m}^{2}$/min   \\
$d_{s}^{(12)}$   & Transition rate of sRNAs towards membrane & $20$ ${\mu m}^{2}$/min   \\
\hline
\end{tabular}
\end{center}
\caption{\label{tab:table-name}Definitions and estimated values of model parameters, adapted from Refs \cite{levine_quantitative_2007,golding_rna_2004,montero_llopis_spatial_2010}.}
\end{table}

\begin{figure}
\begin{flushright}
\includegraphics[width=1.0\columnwidth]{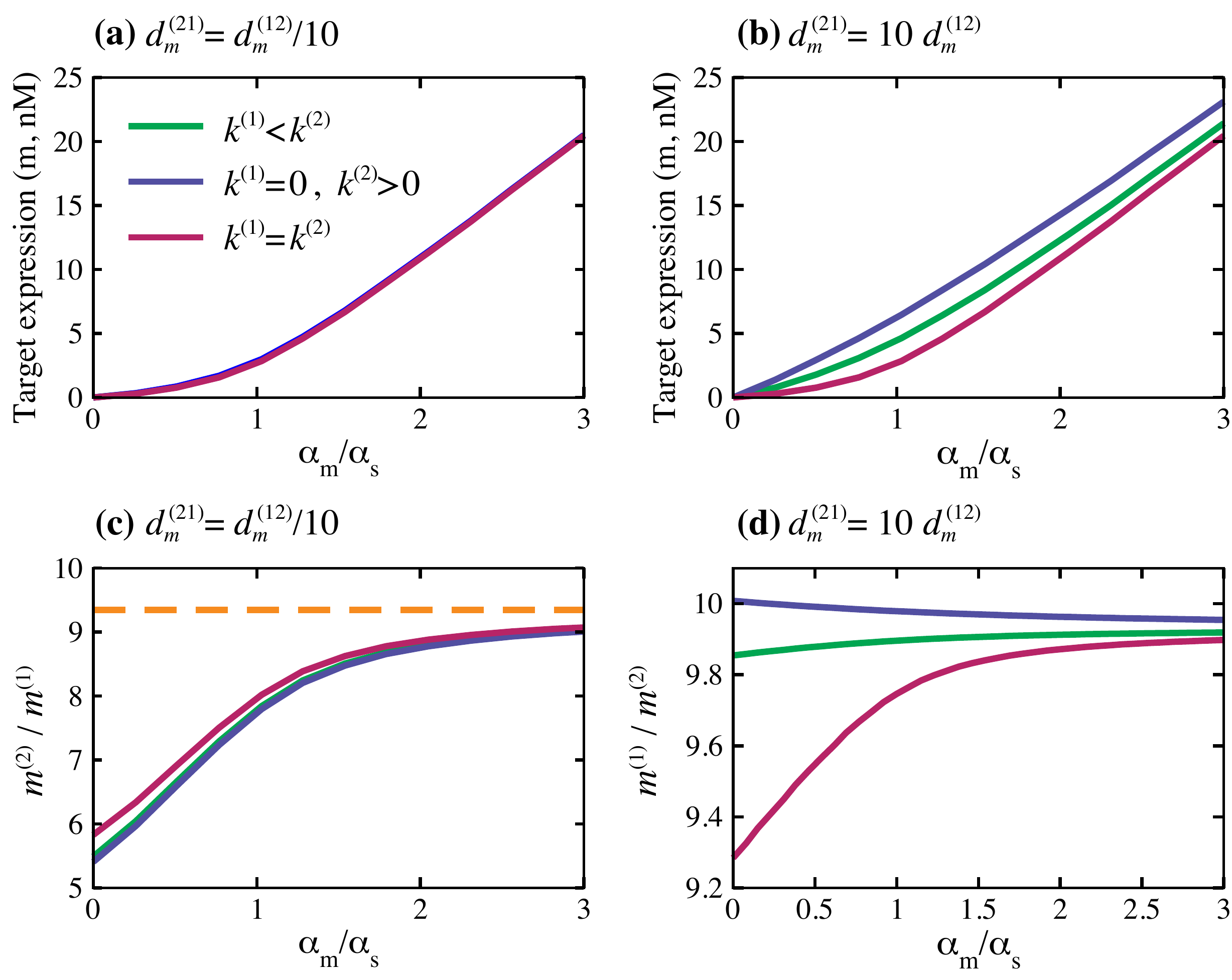}
\end{flushright}
\caption{\textbf{Impact of mRNA localization on the threshold-linear behavior for spatially varying sRNA-mRNA interaction strength}. (a) Total concentration of mRNA for targets biased towards the membrane. (b) Total concentration of mRNA for targets biased away from the membrane.  (c) Enrichment near the membrane for targets biased towards the membrane. (d) Enrichment in region 1 for targets biased away from the membrane. $k^{(2)}=1/50$(nM min)$^{-1}$ and $k^{(1)}=k^{(2)}/5$.  In this and the subsequent figure $R_1 = 0.7 R$. \label{threshold_inh}}  
\end{figure}
        
\section{\label{sec:exact} Spatial dependence of sRNA-mRNA interaction strength}

A prime feature of sRNA-mediated gene regulation is a threshold-linear response, by which target expression is repressed below a threshold value of its transcription rate ($\alpha_{m}<\alpha_{s}$, termed the {repression domain}) and is activated above it   ($\alpha_{m}>\alpha_{s}$, the {expression domain}) \cite{levine_quantitative_2007}. In the spatially-homogeneous case the threshold value is specified by the transcription rate of the sRNA, and the sharpness of the transition from the repressed regime to the activated regime is characterized by the parameter combination $\lambda=\beta_{s}\beta_{m}/k$ termed the leakage rate. 

One of our main goals in this paper is to study the effects of the spatial distribution of sRNA and its target on the threshold-linear response. In this section we focus on a possibility that the spatial distributions of co-factors in the sRNA pathway imply spatial heterogeneity on the kinetics of sRNA-mRNA interactions. This spatially varying sRNA-mRNA interaction strength, in turn, can affect the spatial pattern of the sRNA and its target.

In enterobacteria, two co-factors of the sRNA pathway are involved in the interaction of an sRNA and its target\cite{morita_rnase_2005}. The RNA chaperone Hfq catalyzes pairing of sRNA and mRNA\cite{vogel_hfq_2011}, and duplex formation proceeds slower without Hfq\cite{kawamoto_base-pairing_2006}.
The degradosome endonuclease RNase E is essential for degradation of the sRNA-mRNA duplex \cite{mackie2013rnase}. Recent experiments suggest that in \textit{E. coli} Hfq is enriched near the membrane \cite{diestra2009cellular} and confirm that  RNase E is membrane-bound \cite{khemici_rnase_2008,liou2001rna,taghbalout2007rnasee}.

In the supporting information we show how these spatial distributions can be incorporated in our model as a spatially varying sRNA-target interaction strength $k$. Higher concentration of Hfq and RNase E at the membrane is manifested in our model as a stronger sRNA-mRNA interaction  in region 2, $k^{(2)}>k^{(1)}$. Before addressing some realistic scenarios, it is useful to note that the extreme case $k^{(1)}=0$ lends itself to an exact solution (see supporting information for details). In this case the mean mRNA concentration is given by 
\begin{eqnarray}\label{new_mass_action}
\hspace{-25mm}m=\frac{\gamma}{2\beta_{m}}\left[(A'\alpha_{m} - B\alpha_{s}-\lambda') + \sqrt{(A\alpha_{m} - B\alpha_{s}-\lambda')^{2} + 4\alpha_{m}A\lambda'} \right] \,,
\end{eqnarray}
with 
$A=[1 + \ell_{m1}^2/V_1 + \ell_{m1}^2/V_2]/[(1+\ell_{m1}^2/V_1],\, 
A'=[3 + \ell_{m1}^2/V_1 + \ell_{m1}^2/V_2]/[(1+\ell_{m1}^2/V_1],\,
\mbox{ and }\lambda=\beta_{m}\beta_{s}/k^{(2)}\,.$ Here $\ell_{m1}=\sqrt{d^{(12)}/\beta_m}$ is a  length scale associated with mRNA diffusion from region 1. 
%
%

This expression for $m$ has the same form as the threshold-linear solution of the mass-action model of a spatially homogeneous system with unbiased kinetics. In the inhomogeneous system the threshold is shifted from $\alpha_m = \alpha_s$  to 
\begin{equation}\label{inho_th}
\alpha_m = 
\frac{\gamma + \ell^2_{s1}/V_1}{1 + \ell^2_{s1}/V_1}\alpha_{s}
\end{equation} 
with $\ell_{s1}=\sqrt{d_s^{(12)}/\beta_s}$.
Notably, the transition point is still completely determined by the kinetic properties of the sRNA. In the typical case of fast sRNA diffusion $\ell_{s1}^2/V1$ is very large (it is $\gtrsim 1000$ with the parameters of Table 1), and  the transition occurs at  $\alpha_m \simeq \alpha_s$. 
The sensitivity at the transition is governed by the 
renormalized leakage rate 
\begin{equation}\label{inho_leakage}
\lambda'=\lambda\;\frac{1+\ell_{m1}^2/V_1+\ell_{m2}^2/V_2}{1+\ell_{m1}^2/V_1}\;\frac{1+\ell_{s1}^2/V_1+\ell_{s2}^2/V_2}{1+\ell_{s1}^2/V_1}\,.
\end{equation}
Inspection of this expression suggests that the inhomogeneity in $k$ can only increase the smoothness of the transition. If diffusion is fast, then such an increase is only significant if either the mRNA or the sRNA is biased away from the membrane,  
$\ell_{m2}^2/V_2 \gg \ell_{m1}^2/V_1$ or $\ell_{s2}^2/V_2 \gg \ell_{s1}^2/V_1$.  The former case may be a model for genes whose mRNAs are localized to the vicinity of their transcription site \cite{nevo2011translation,buskila_rna_2014}. 
Under these conditions  mRNAs are biased away from the region of interaction with the sRNA,  allowing them to escape the sRNA-mediated degradation.

What lessons can we learn from this analysis of the extreme case $k^{(1)}=0$? First, we verified numerically that all the conclusions drawn for this case also hold for a finite $k^{(1)}<k^{(2)}$ (Fig.~\ref{threshold_inh}(a-b)). The enhanced coupling with the sRNA near the membrane can bias the mRNA distribution in the cell, although not by much (Fig.~\ref{threshold_inh}(c-d)).  Second,  these results led us to conjecture that the effect of spatial heterogeneity in the cell can   be completely captured by a simple spatially homogeneous model through renormalization of the leakage rate $\lambda$. Below we test this conjecture numerically and use it to analyze implications  of spatial heterogeneity. 

%
%

\begin{figure}[ht]
\begin{flushright}
\includegraphics[width=1.0\columnwidth]{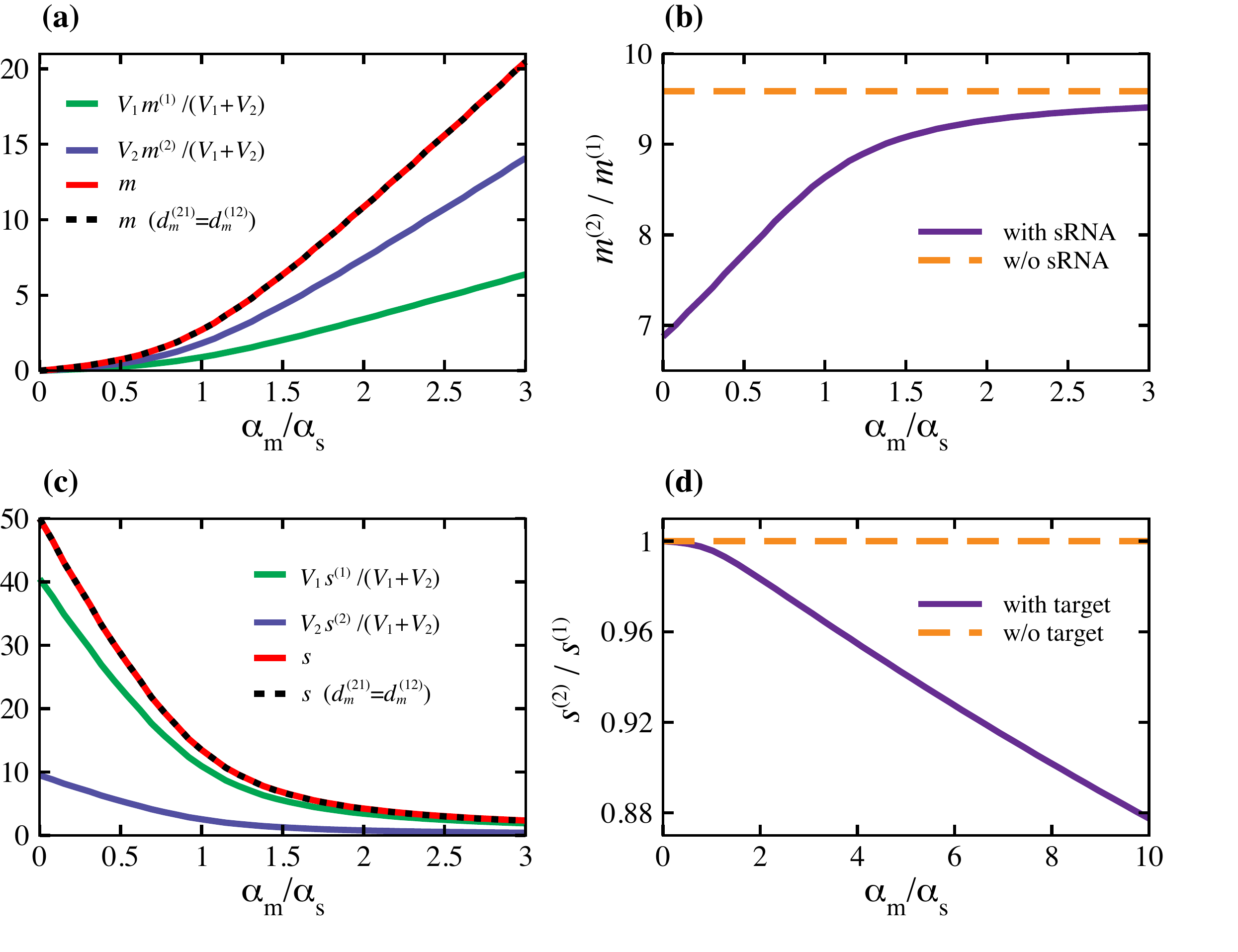}
\end{flushright}
\caption{\textbf{Response of membrane-enriched target to an unbiased sRNA.} 
(a) Total mRNA concentration and the fraction in each region for a membrane-enriched target ($d^{(21)}=d^{(12)}/10$) at different levels of expression, compared with an unbiased target. In all figures of Sections 3 and 4  $R_1 = 0.9 R$. 
(b) Effect of the sRNA on enrichment near the membrane. 
(c) Total sRNA concentration and the fraction in each region in the presence of membrane-enriched target at different levels of expression, compared with an unbiased target. 
(d) Effect of the interaction with the membrane-enriched target on sRNA distribution. 
\label{mRNA_localization}} 
\end{figure}
\begin{figure}
\begin{flushright}
\includegraphics[width=1.0\columnwidth]{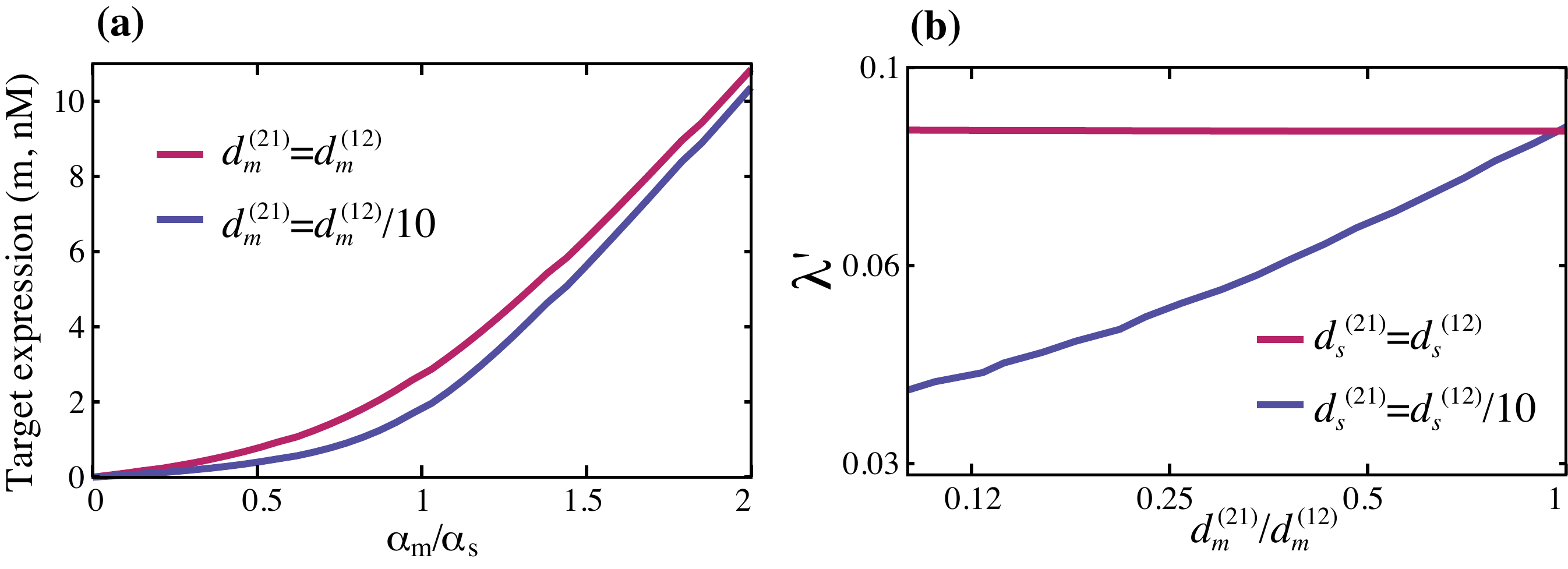}
\end{flushright}
%
\caption{\textbf{Response of mRNA targets to membrane-enriched sRNA.} 
(a) Total mRNA concentration of a membrane-enriched and an unbiased target in the presence of a membrane-biased sRNA ($d_s^{(21)}=d_s^{(12)}/10$) at different levels of expression. 
(b) The effect of mRNA localization on the renormalized leakage rate in the presence of biased or unbiased sRNA. \label{thresholdlinear}} 
\end{figure}

\section{Responses of mRNA targets to biased and unbiased sRNAs}
The bias of mRNAs  that encode membrane-associated proteins towards the membrane may affect their regulation by sRNA \cite{kawamoto_implication_2005,fei_determination_2015}. 
In other cases,  enrichment near or at the membrane of RNA-binding proteins (like Hfq \cite{diestra2009cellular})  may lead to bias in the mobility of associated sRNAs. In this section we study the effect of these kinetic biases on sRNA regulation. 

With an unbiased sRNA the threshold-linear response of a membrane-enriched mRNA is identical to that of a uniformly distributed target (Fig.~\ref{mRNA_localization}a,c). Enrichment near the membrane, quantified here by the ratio $m^{(2)}/m^{(1)}$, is up to 40\% lower in the repression domain ($\alpha_m<\alpha_s$) than in the expression one ($\alpha_m>\alpha_s$, Fig.~\ref{mRNA_localization}b). In contrast, the sRNA is homogeneously distributed in the repression domain, as expected from its fast unbiased diffusion. However, in the expression domain, when the mRNA becomes more abundant and highly biased towards region 2, more sRNAs are lost due to coupled degradation in that region. With high mRNA expression, this effect cannot be washed out by the fast diffusion, and the sRNA becomes depleted near the membrane (Fig.~\ref{mRNA_localization}d). At the high end of attainable  transcription rates $s^{(2)}/s^{(1)}$ can reach 0.75 and below (not shown).

To see how the spatial bias of both the sRNA and its target affects the threshold-linear response, we plotted the total concentration of target mRNA as a function of its transcription rate in Fig.~\ref{thresholdlinear}(a). It can be seen that the bias of mRNAs towards the membrane increases the sharpness  of the transition from the repression domain to the expression domain (the `sensitivity'). 
To make this observation more quantitative we utilize our results from section \ref{sec:exact}, which suggest that the steady state of the inhomogeneous system can be described as the steady state of a homogeneous system upon renormalization of its parameters. Estimation of the rescaled leakage rate $\lambda'$ is done by first solving the full model numerically, and fitting the resulting  $m(\alpha_m)$ curve to a solution of the homogeneous model (see supporting information).  In Fig.~\ref{thresholdlinear}(b) we plotted   $\lambda'$ as a function of the rate of release of the target mRNA from the membrane, $d_{m}^{(21)}$. When the sRNA exhibits unbiased diffusion the leakage rate is insensitive to spatial distribution of mRNA (red curves), but when the sRNA is enriched near the membrane the leakage rate decreases when the bias of the mRNA towards the membrane increases (blue curve). This is the result of co-localization of sRNA and mRNA, which facilitates the stoichiometric sRNA-mRNA degradation, lowers the leakage rate, and increases the sharpness of the transition. These observations demonstrate that the spatial localization can control the regulation of gene expression by effectively tuning the coupled degradation of the sRNA and its target.

\section{The role of spatial distribution on the hierarchy and cross-talk between  targets}

It has been shown that the interaction strengths between an sRNA and its targets determine a hierarchy among the targets, such that strongly interacting targets are repressed before weakly interacting ones \cite{levine_quantitative_2007,fei_determination_2015,mitarai_efficient_2007}. 
To investigate the role of spatial localization in influencing the hierarchy we consider two targets of the same sRNA, one of which is biased towards the cell membrane (target species 1) while the other is uniformly distributed in the cell (target species 2). We assume that the two targets are otherwise described by equal reaction rates indicated in Table~1. 

\begin{figure}
\begin{flushright}
\includegraphics[width=1.0\columnwidth]{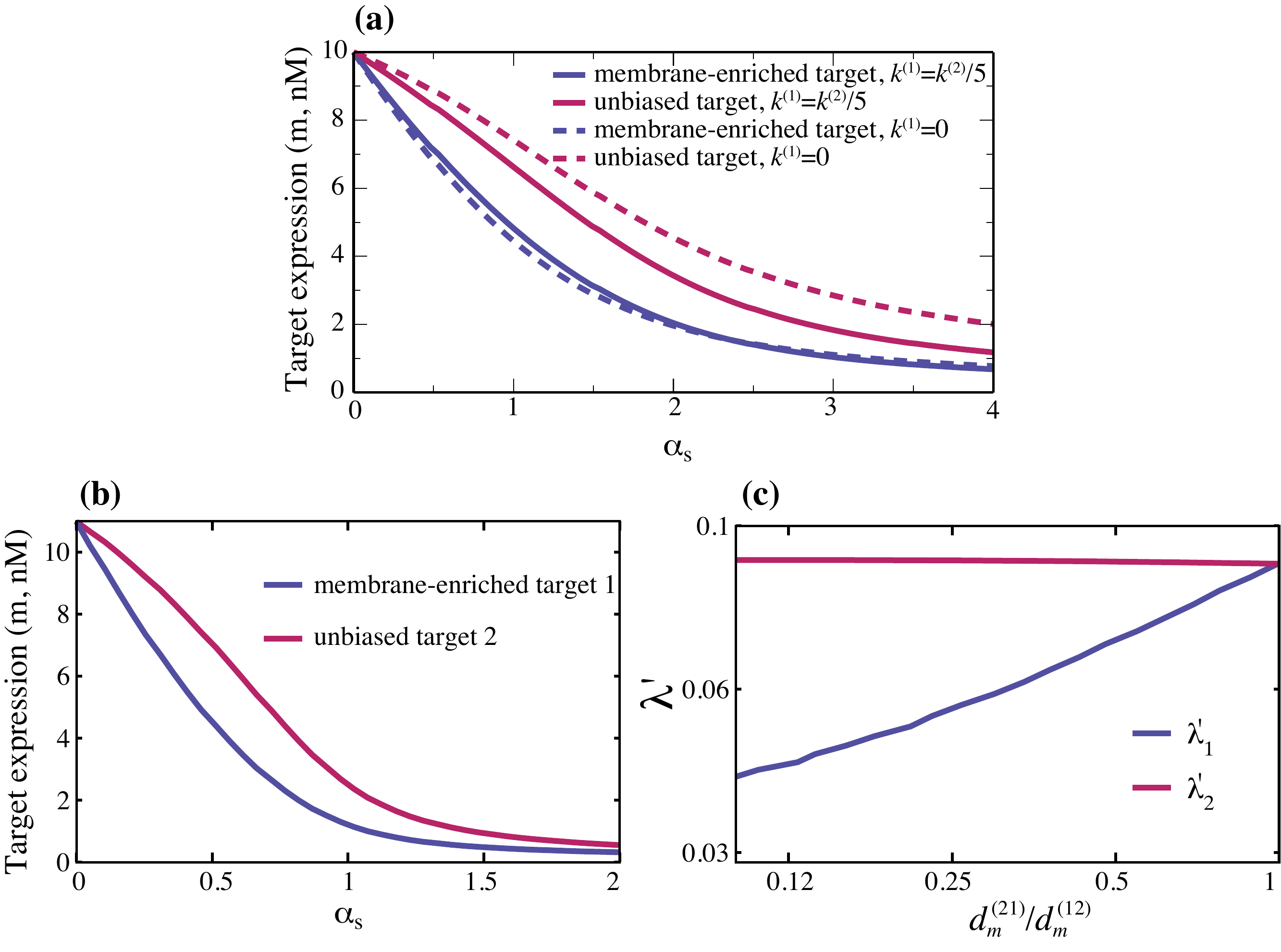}
\end{flushright}
\caption{\textbf{Hierarchy of targets as a result of localization.} 
(a) Hierarchy between two targets in the case of spatially varying sRNA-target interaction strength.
(b) Concentration of two targets of a membrane-biased sRNA as a function of the transcription rate of the sRNA. 
(c) The renormalized leakage rate of both targets of panel (b). \label{fig:hierarchy}} 
\end{figure}

When all parameters describing the sRNA kinetics are spatially unbiased, the two targets respond equally to the sRNA (not shown). This is not the case when the sRNA experiences biased kinetics. As described above, spatial heterogeneities of sRNA co-factors can be taken into account in two ways, with similar results: by having $k^{(1)}<k^{(2)}$ or by biasing the sRNA mobility $d^{(21)}<d^{(12)}$. As with a single target, one can again solve the limit $k^{(1)}=0$ exactly, confirming that even with multiple targets the inhomogeneous system can be described in terms of a homogeneous system with renormalized parameters (see supporting information). Taking this approach, we find that spatial bias in the sRNA -- regardless of the mechanism -- can lead to a hierarchy between the two targets: the target that is biased towards the membrane is suppressed more strongly and at lower values of $\alpha_s$ than the uniformly distributed one (Fig.~\ref{fig:hierarchy}(a,b)). This can be attributed to a reduction in the effective leakage rate  $\lambda'$ of the biased target, as the leakage rate of the unbiased target remains unchanged  (Fig.~\ref{fig:hierarchy}(c)).

\begin{figure}
\begin{center}
\includegraphics[width=0.5\columnwidth]{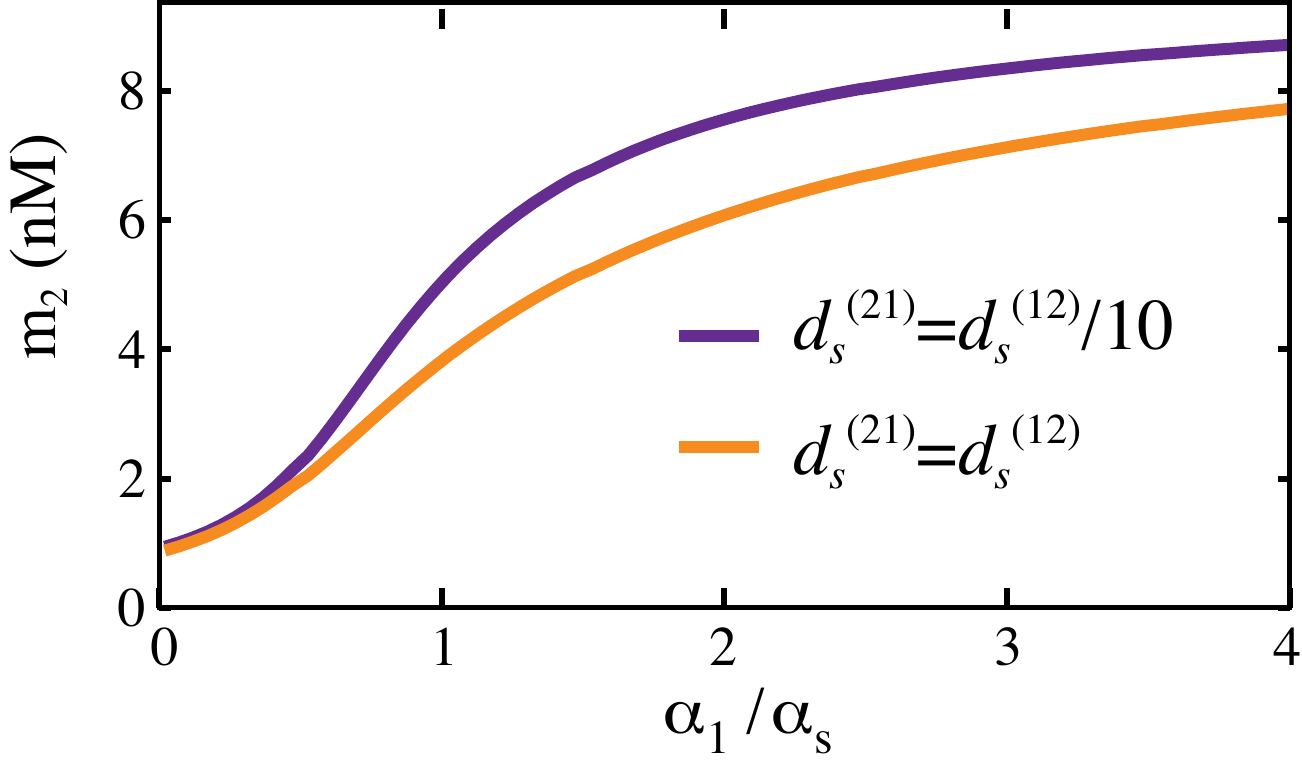}
\vspace{-12pt}
\end{center}
\caption{\textbf{Cross-talk between two targets.} Concentration of unbiased target 2 as a function of the transcription rate of membrane-enriched target 1. Here   $d_{1}^{(21)}=d_{1}^{(12)}/10$, $d_{2}^{(21)}=d_{2}^{(12)}$, $\alpha_{s}=2$ nM/min and $\alpha_2 = 1$ nM/min.\label{cross_talk}} 
\end{figure}

Since each target reduces the sRNA concentration, the expression of one target can influence the expression of the other target \cite{levine_quantitative_2007,feng_qrr_2015}. As the strength of this cross-talk depends on the coupled degradation rate of each target with sRNA, it is natural to ask how localization influences it. In Fig.~\ref{cross_talk} the cross-talk is quantified through the change in concentration of the unbiased target~2 on the transcription rate of membrane-enriched target~1. As in the spatially homogeneous case, target 2 is repressed when $\alpha_{1}+\alpha_2<\alpha_{s}$ and activated for $\alpha_{1}+\alpha_2>\alpha_{s}$. As seen in Fig.~\ref{cross_talk}, co-localization of the sRNA and target 1  increases the level of cross-talk. In this case, co-degradation between a localized target and a biased sRNA efficiently eliminates the sRNA, allowing the second target to be expressed.

Since spatial bias of a target mRNA can induce spatial bias on the sRNA (Fig.~\ref{mRNA_localization}d) we asked if the cross-talk between targets can also induce a spatial bias on an otherwise homogeneously distributed target. Our results show no support of this possibility, due to the fact that the spatial bias induced on the sRNA by its target is only significant when the level of sRNA is highly suppressed.

\section{Summary and discussion}
Since diffusion in the bacterial cell is very fast in view of its small dimensions, most theoretical studies of genetic circuits in bacteria -- both natural and synthetic -- ignore any spatial aspect, and assume that the cell content is well mixed. Recent findings that the transcript of some gene families display non-trivial spatial structure raise the question whether analysis of regulatory circuits that employ post-translational regulation should consider these spatial aspects in more detail.

Here we developed a minimal theoretical framework for addressing this question in the context of small regulatory RNAs. Our results predict that the spatial localization of mRNA can modulate quantitative  characteristics of sRNA-mediated regulation, including the position and smoothness of a threshold-linear response typical of such systems. In addition, our results suggest that the spatial distribution of two targets modulates the established hierarchy and the cross-talk among  targets that share the same sRNA. Importantly, we find that these changes are only noticeable when the sRNA itself is inherently biased, either through its translocation dynamics or through its association with co-factors that are unevenly distributed in the cell. Even in these cases, fast diffusion in the cell guarantees that the relative effect is never very large. 

The small RNA SgrS is a regulator of the Glucose PTS transporter gene \textit{ptsG} in \textit{E. coli} \cite{vanderpool_involvement_2004,kawamoto_implication_2005,fei_determination_2015}. Through genetic manipulation it has been shown that the efficiency of repression of \textit{ptsG} expression by SgrS depends on the localization of ptsG transcripts to the vicinity of the membrane \cite{kawamoto_implication_2005}. This has been suggested to be the effect of enrichment of ribosomes near the membrane. Our results suggest that the spatial bias of Hfq and RNase E may be alternative of additional contributes to the mechanisms behind this observation. 

The spatial distribution of target mRNAs may impact the functionality of genetic circuits, both natural and synthetic. This is particularly true  when multiple genes within a circuit are regulated by the same sRNA. In such cases, the differential localization of the targets may set the hierarchy among these genes, which could affect the logic and the temporal dynamics of the circuit. Synthetic biological circuits which mix membrane-bound components with cytoplasmic ones should be designed with care for such prioritization and kinetic effects.  

Transcriptional regulation of gene expression is an efficient mechanism in which binding at a single genomic locus can have significant regulatory impact. In contrast, post-transcriptional regulation requires direct interaction with multiple target molecules. This gives rise to stoichiometric effects and is the basis of the threshold-linear response. Moreover, this raises the possibility that under some condition molecules of one gene would compete with molecules of another gene  for a common regulator (such as small RNA) or for auxiliary co-factors (such as Hfq \cite{Moreno_2015}). This competition can modulate the impact of the regulator on each one of its targets. Combined with fact that  most sRNAs have multiple targets, this raises the hypothesis that some RNAs in the cell only interact with an sRNA to modulate its interactions with other targets.   These are known as competing endogenous RNAs (ceRNAs) \cite{Salmena_2011,jost_small_2011,jost2013regulating,bossi2016competing}. In our model we considered the case of two competing targets, and found that their spatial organization affects their level of cross-talk only in the context of spatially biased sRNA. While in multi-cellular organisms the co-expression of targets in the same tissues is critical for their competitive effect \cite{Guttman_2012}, no such requirement exists for intra-cellular co-localization.


In this paper we focused on two types of sub-cellular distributions that roughly follow the longitudinal symmetry of the cell. We mostly focus on genes that express membrane-bound proteins, whose mRNAs are known to be enriched near the membrane, and briefly consider targets  whose mRNA has some affinity to the nucleoid. Important spatial patterns that do not follow this symmetry are  enrichment  of proteins and their mRNA near the poles of the cell \cite{nevo2011translation,buskila_rna_2014,castellana_spatial_2016} and localization to the nascent septum separating daughter cells \cite{dos2012diviva}. While our model does not directly apply to these localization patterns, we expect such mRNA to show similar response to sRNA regulation as the ones discussed here. 

In bacterial gene expression, one main source of intrinsic stochasticity is the effect of rare transcription events which leads to a burst of proteins. It has been suggested that sRNA may impede intrinsic noise by  decreasing  the protein burst size by reducing the translation rate and by decreasing the stability of the mRNA\cite{levine2008small,mehta_quantitative_2008}. As discussed in the text, under some conditions spatial bias of both sRNAs and their targets can effectively increase the  strength of the interaction between an sRNA and its target, in addition to reducing their available space for free diffusion. Under such conditions it is expected that qualities of an sRNA as a noise suppressor could be enhanced.

The natural next step would be to test these predictions experimentally in order to increase our understanding of roles of spatial organization in bacteria. One possible way to investigate the role of cofactors of the sRNA pathway on the spatial localization is to inhibit the activity of RNase E \cite{morita2011rnase, masse2003coupled} or use functional mutants of Hfq \cite{panja2015acidic}. Further theoretical and experimental investigations are required to reveal the roles of spatial organization in the regulation of bacterial physiology and metabolism.

\section*{Acknowledgments}
We thank Neil Peterman and Rinat Arbel-Goren for discussions. This work was supported in part by the NSF  grant MCB-1413134 (EL) and by the ISF grant 514415 (JS). JS is the incumbent of the Siegfried and Irma Ullman Professorial Chair.

\section*{References}

\bibliography{bibfile}
\bibliographystyle{plos2015}

\end{document}